\documentclass[prc,aps,floatfix,nofootinbib,twocolumn,superscriptaddress]{revtex4} 
\usepackage{graphicx} 
\usepackage{amssymb} 
\usepackage{amsmath} 
\usepackage{amsfonts} 
\usepackage{bbold}
\usepackage{color}

\renewcommand{\vec}[1]{\mbox{\boldmath $#1$}}

\def\be{\begin{equation}} 
\def\ee{\end{equation}}

\begin{document} 

\title{
Microscopic derivation of transition-state theory for complex quantum systems
}

\author{Kouichi Hagino}
\affiliation{ 
Department of Physics, Kyoto University, Kyoto 606-8502,  Japan} 

\author{George F. Bertsch}
\affiliation{ 
Department of Physics and Institute for Nuclear Theory, Box 351560, 
University of Washington, Seattle, Washington 98915, USA}

\begin{abstract}
The decay of quantum complex systems through a potential barrier is often
described with transition-state theory, also known as RRKM theory
in chemistry.  Here we derive the basic formula for transition-state theory
based on a generic Hamiltonian as might be constructed in a
configuration-interaction basis.   
Two reservoirs of random Hamiltonians from Gaussian orthogonal ensembles are coupled to
intermediate states representing the transition states at a barrier.  
Under the condition that the decay of the reservoirs to open channels
is large, an analytic formula for reaction rates is derived.
The transition states act as independent Breit-Wigner
resonances which contribute additively to the total transition probability,
as is well known for electronic conductance through resonant tunneling states.
It is also found that the
transition probability 
is independent of the decay properties of the states in the second reservoir 
over a wide range of decay widths.
\end{abstract}
 
\maketitle

\section{Introduction}

Transition-state theory is ubiquitous in physics and chemistry to calculate
reaction and decay rates for many-particle systems in the presence of a barrier
\cite{hanggi1990,tr96,weiss2012,BW39,MR51}.
The assumptions in the theory are clear in classical dynamics but less so
in the quantum regime.  For fermionic systems of equal-mass particles, the Hamiltonian is often
formulated in a configuration-interaction (CI) representation.  This motivates
considering models that exhibit the barrier dynamics in the CI framework
to understand conditions to support transition-state approximations. 

\section{Model Hamiltonian}
Following previous recent work, we consider here a  Hamiltonian
composed of three sets of states. The states in the barrier region are
represented in the Hamiltonian $H_2$.  Their precise structure is not
specified, but we have in mind a set of configurations with the ground
states and their quasiparticle excitations determined by constrained Hartree-Fock or 
density-functional theory.  The other two sets of states
contained in Hamiltonians $H_1$ and $H_2$ are statistical reservoirs,
with their Hamiltonians constructed from the matrices of the Gaussian
orthogonal ensemble GOE \cite{weidenmuller2009}.  
For this Hamiltonian, one may think about e.g., a decay from a highly excited 
configuration, so that the both the pre-saddle and the post-saddle configurations 
can be treated 
statistically \cite{bertsch-hagino2023,polik1990,miller1990,hernandez1993}. 
It is important to note that the GOE
Ansatz is the only statistical input, and the ensemble  is microcanonical
rather than canonical.  The full Hamiltonian 
reads
\begin{equation}
H=
\left(
\begin{matrix} 
H_1 & V_{12} & 0 \cr
V_{12}^T & H_2 & V_{32}^T \cr
0 & V_{32} & H_3
\end{matrix}
\right), 
\label{eq:H}
\end{equation}
where $V_{12}$ and $V_{23}$ are matrices of the coupling interaction between
the reservoir states and the states in the bridge Hamiltonian.  This model
is generalization of the 
of the model in Ref. \cite{weidenmuller2022}, which assumes that $H_2$ has
a single state at the barrier top.
See also Ref. \cite{weidenmuller2023} for a similar generalization. 

\begin{figure}[tb] 
\begin{center} 
\includegraphics[width=\columnwidth]{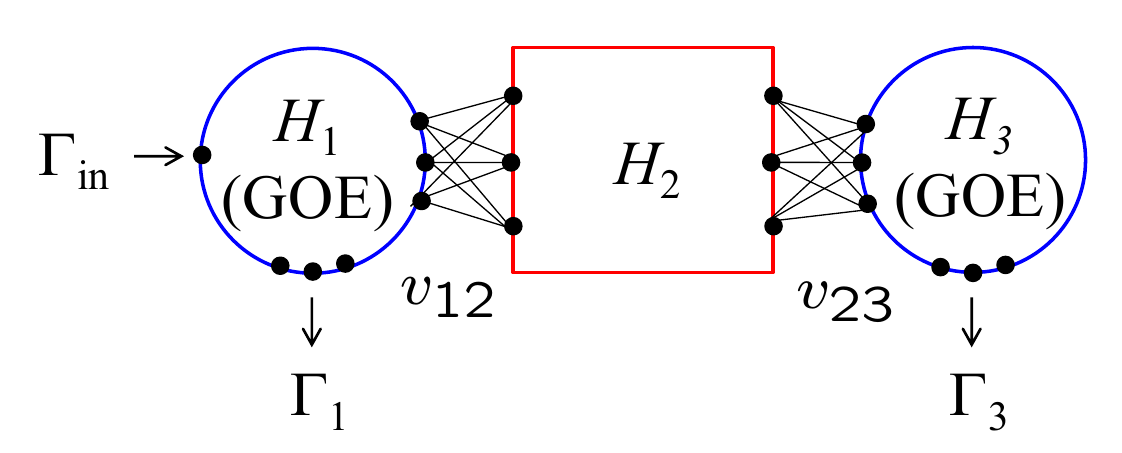} 
\caption{Schematic structure of the model Hamiltonian composed
of two reservoirs connected by a bridge Hamiltonian and open to
three sets of external channels.
}
\label{fig:1} 
\end{center} 
\end{figure}

To complete a model for reactions, one also needs the coupling
matrix elements between $H$ and the  reaction
channels.  With those ingredients the
$S$-matrix for transitions from one channel to another can be computed
by standard linear algebra manipulations.  If one is only interested in 
reaction probabilities,  the linear algebra can be collapsed to a compact
formula \footnote{The formula had been used earlier
as well \cite{meir1992} in the theory of electrical conductivity. }
\cite{mil93,da01,da95,ha08,al21} for the transition probability from
channel $a$ to channel $b$ given by
\be
T_{ab} = | S_{ab}(E)|^2 = {\rm Tr}\left( \Gamma_a G(E) 
\Gamma_b G^{\dagger}(E)\right).
\label{eq:datta}
\ee
Here 
$G$ is the Green's function of the Hamiltonian in presence of 
entrance or decay channels $c$ at reaction energy $E$,
\footnote{
It is implicitly assumed in 
Eq. (\ref{eq:GG}) that the $\Gamma_c$ are independent of energy.}
\be
G(E) = \left(H -i \sum_c \Gamma_c/2 -E\right)^{-1}.
\label{eq:GG}
\ee
In general the  $\Gamma_c$ are rank-one matrices (of the same dimension as
$H$)  but for the present model they have only one entry on the  diagonal
and  a block structure given by
\begin{equation}
\Gamma_{\rm in}=\left(
\begin{matrix} 
\tilde\Gamma_{\rm in} & 0 & 0\cr
0 & 0 & 0 \cr
0 & 0 & 0
\end{matrix}
\right), \,
%
\Gamma_1=\left(
\begin{matrix} 
\tilde\Gamma_{1} & 0 & 0\cr
0 & 0 & 0 \cr
0 & 0 & 0
\end{matrix}
\right), \,
\Gamma_3=\left(
\begin{matrix} 
0 & 0 & 0\cr
0 & 0 & 0 \cr
0 & 0 & \tilde\Gamma_{3}
\end{matrix}\right)
\label{eq:Gin}
\end{equation} 
depending on which subblock the channel $c$ connects to.
The full Hamiltonian with its coupling to external channels is
depicted in Fig. \ref{fig:1}.
In general, one is
interested in the reaction probability $T_{a3}$ from an entrance channel
($a={\rm in}$) to all possible decay channels in the second reservoir,
\be
T_{{\rm in},3} = \sum_{c \in {\rm block}\,\,\, 3} |S_{{\rm in}, c}|^2.
\ee
Due to the block structures of $H$ and $\Gamma_c$ we only need the $G_{13}$
block of the Green's function
\begin{equation}	
G=
\left(
\begin{matrix} 
G_{11} & G_{12} & G_{13} \cr
G_{21} & G_{22} & G_{23} \cr
G_{31} & G_{32} & G_{33} 
\end{matrix}
\right) 
\label{eq:G}
\end{equation}
in Eq. (\ref{eq:datta}). 
As derived in Appendix A, the submatrix $G_{13}$ reduces to 
\begin{equation}
G_{13}=G_1V_{12}G_2V_{32}^TG_3,
\label{eq:G13}
\end{equation}
where $G_1$, $G_3$, and $G_2$ are given by
\begin{eqnarray}
G_1&=&\left(H_1-i\tilde\Gamma_{\rm in}/2
-i\tilde\Gamma_{1}/2-E\right)^{-1},  
\label{eq:G1-1}
\\
G_3&=&\left(H_3-i\tilde\Gamma_{3}/2-E\right)^{-1}, 
\\
G_2&=&\left(H_2-V_{12}^TG_1V_{12}-V_{32}^TG_3V_{32}-E\right)^{-1}.
\label{eq:G2-3}
\end{eqnarray}
Substituting Eq. (\ref{eq:G13}) into Eq. (\ref{eq:datta}), the 
transmission coefficient is obtained as
\begin{eqnarray}
T_{\rm in,3}(E)
&=&{\rm Tr}[
\tilde\Gamma_{{\rm in}}(G_1V_{12}G_2
V_{32}^TG_3)\tilde\Gamma_{3}(G_3^\dagger V_{32}G_2^\dagger V_{12}^TG_1^\dagger)], 
\nonumber \\
\\
&=&{\rm Tr}[
(V_{12}^TG_1^\dagger \tilde\Gamma_{\rm in}G_1V_{12})G_2
(V_{23}G_3\tilde\Gamma_{3}G_3^\dagger V_{32})G_2^\dagger]. \nonumber \\
\label{eq:datta2}
\end{eqnarray}
We write the elements of the two GOE
Hamiltonians as
\begin{equation}
(H_k)_{ij}=(H_k)_{ji}=v_{k}\sqrt{1+\delta_{ij}}\,r_{ijk}, 
\end{equation}
where $r_{ijk}$ is a random number from a Gaussian distribution of unit dispersion, 
$\langle r_{ijk}^2\rangle=1$.
Then the average level density $\rho_k$ at $E=0$ at the centers of the GOE Hamiltonians
is given by
\be
\rho_k = \frac{N_k^{1/2}}{\pi v_k}, 
\ee
where $N_k$ is the dimension of $H_k$.
We set $E=0$ for the rest of this paper.
Each state in $H_2$ is assumed to couple to specific states in the
GOE reservoirs.  We parameterize the couplings as \footnote{We 
implicitly assume $N_k> N_2$ for both reservoirs.}
\begin{equation}
(V_{12})_{ij}=v_{12}N_1^{1/2}\delta_{ij},~~~(V_{32})_{ij}=v_{32}
N_3^{1/2}\delta_{ij}. 
\end{equation}
This parameterization is not as restrictive as it may seem.  Due to
the GOE invariance, the couplings can be to arbitrary orthogonal
vectors in the GOE spaces.  The specific form of the coupling
is such that the average matrix element is independent of the
dimension $N_k$.  

The matrices for the decay widths are assumed to be diagonal with elements 
\begin{equation}
(\tilde\Gamma_1)_{ij}=\gamma_1\,\delta_{i,j},~~~(\tilde\Gamma_3)_{ij}=\gamma_3\,\delta_{i,j}, 
\end{equation}
except for the entrance channel $a={\rm in}$, which couples to a single
state $ i = 1$ in the first reservoir,
\begin{equation}
(\tilde\Gamma_{\rm in})_{ij}=\gamma_{\rm in}\,\delta_{i,1}\delta_{j,1}. 
\label{eq:Gin2}
\end{equation}
Without loss of generality, the only requirement on $\tilde \Gamma_{\rm in}$ is that it
has rank one within the space of $H_1$, as in the  ``B\"uttiker
probes" of semiconductor transport theory\cite{buttiker1986,venugopal2003,datta2005}.

\section{Transition-state theory}

We now examine how the average reaction probability depends on the
parameters of the model.  
Since transition-state theory deals with fluxes into or from a statistical
reservoir, it is convenient to define a transmission coefficient $\mathcal{T}_{ck}$ 
of a channel $c$ in the second block into the reservoir $k$
\be
\mathcal{T}_{ck} = 2 \pi \rho_k \gamma_k 
\ee
and its sum  over channels,
\be
\mathcal{T}_k = \sum_{c \in k} \mathcal{T}_{ck}.
\ee
For small values of $\mathcal{T}_{ck}$ it has the physical significance of the
transmission probability from the channel $c$ into the statistical
reservoir.     
As shown in Ref. \cite{weidenmuller2022}, it is straightforward to carry out the statistical
averaging for Eq. (\ref{eq:datta}) in the limit 
that $\mathcal{T}_k/N_k  \gg 1$
for both reservoirs.  
We first examine the Green's function $G_2$ and the 
coupling terms $V_{k2}^TG_kV_{k2}$ in it.   The averages
of $G_k~(k=1,3)$ at $E=0$ 
\footnote{
We note that there are mild restrictions on the ranges of the parameters in Eq.
(\ref{eq:Gk}).
In practice, the widths associated with the individual channels should
be large compared to the level spacing in the GOE but small with respect to 
the boundaries of its eigenspectrum.} including the decay-width
matrices are given by \cite{weidenmuller2022,Lo2000,Fe2020,hagino-bertsch2021}
\be
\langle G_{k}\rangle_{ij} =i\frac{\pi\rho_k}{N_k} \delta_{ij}.
\label{eq:Gk}
\ee   
The standard deviation of the fluctuations is 
\be
SD(G_{k})_{ij} =\frac{\pi\rho_k}{N_k}(1+i)
\left(\frac{2(1+\delta_{i,j})N_k}{\mathcal{T}_k}\right)^{1/2}.
\label{eq:GSD}
\ee
The fluctuations go to zero in the limit $\mathcal{T}_k/N_k  \gg 1$ so these terms in $G_2$
can be replaced by $i\pi v_{k2}^2 \rho_k$ times the unit matrix.  
Thus 
the correlations between $G_2$ and the other terms in Eq. (\ref{eq:datta2})
vanish, allowing it to be evaluated as
\be
\bar G_2 = (H_2-V_{12}^T\langle G_1\rangle V_{12}-V_{32}^T
\langle G_3\rangle V_{32}-E)^{-1}. 
\ee
The two terms in parentheses in Eq. (\ref{eq:datta2}) are independent of
each other so can also be replaced by their ensemble averages.  
\begin{equation}
\langle (V_{12}^T G_1 \tilde\Gamma_{\rm in}  G_1^\dagger V_{12})_{ij}\rangle
=\frac{\gamma_{\rm in}}{N_1 \gamma_1} 2\pi v_{12}^2\rho_1\delta_{ij} 
\end{equation}
\begin{equation}
\langle (V_{32}^T G_3 \tilde\Gamma_3  G_3^\dagger V_{32})_{ij}\rangle
= 2\pi v_{32}^2\rho_3\delta_{ij}. 
\end{equation}
in the limit $\mathcal{T}_k/N_k \gg 1$ (see Appendix B).  

We can cast the formulas in a more
transparent notation by defining decay widths
of the transition states to the right-hand and left-hand reservoirs
as
\be
\Gamma_R = 2 \pi v_{32}^2\rho_3  
\ee
and
\be
\Gamma_L = 2 \pi v_{12}^2\rho_1.
\ee
Using Eq. (\ref{eq:VG_inGV}) in Appendix B, 
one obtains 
\begin{eqnarray}
\langle T_{\rm in,3}\rangle&=&
\frac{\mathcal{T}_{\rm in}}{\mathcal{T}_1}\,
\Gamma_L\Gamma_R\sum_{i,j}\langle |(G_2)_{ij}|^2\rangle\,
\sum_{b\in3}\left(\frac{T_b}{\sum_{b'\in3}T_{b'}}\right), \nonumber \\
\\
&=&
\frac{\mathcal{T}_{\rm in}}{\mathcal{T}_1}\,
\Gamma_L\Gamma_R\sum_{i,j}\langle |(G_2)_{ij}|^2\rangle 
\label{eq:T_in3-2}
\end{eqnarray}
where 
$\mathcal{T}_{\rm in}=2\pi\Gamma_{\rm in}\rho_1/N_1$. 
It is remarkable that the ensemble average of the transmission coefficient is
independent of $\Gamma_3$, and thus the insensitive property 
\cite{bertsch-hagino2021,hagino-bertsch2021,bertsch-hagino2023} is
realized.

Notice that $G_2$ in Eq. (\ref{eq:G2-3}) can be written 
\begin{equation}
    G_2=(H_2-i(\Gamma_L/2+\Gamma_R/2)\mathbb{1})^{-1}.
\end{equation}
Then if $H_2$ is diagonal\footnote{In fact, this restriction is not
necessary.  The matrix $H_2$ can always be diagonalized by an orthogonal
transformation which has no effect on the ensemble average.} 
with matrix elements $(H_2)_{ij}=E_i\delta_{i,j}$, 
the transmission coefficient becomes
\begin{equation}
\langle T_{\rm in,3}\rangle
=\frac{\mathcal{T}_{\rm in}}{\mathcal{T}_1}\sum_i \frac{\Gamma_L\Gamma_R}{E_i^2+(\Gamma_L+\Gamma_R)^2/4}.
\label{eq:resonances}
\end{equation}
This is a well-known formula for electron transport 
through intermediate resonances \cite{alhassid2000,bertsch2014}. 
It agrees with an underlying assumption in transition-state models,
that the contributions of the individual transition states are
additive in the total transmission probability
\footnote{This may not be the case for the probability fluxes
through the individual transition states.}.  
  
The formula also shows that the contribution to the transmission coefficient 
is suppressed  when the energy of a bridge state is outside the range of $\pm(\Gamma_L+\Gamma_R)/2$ around the incident 
energy.  This is marked contrast to models in which the
transition state is an internal channel that remains open at all
energies above the threshold.  To maintain the correspondence to the
CI formulation, one would have to include highly excited
configurations that carry momentum along a collective coordinate.
At some point the model would break down because the coupling matrix
element would become small compared to $v_k$, the coupling strength
of the configurations within the GOE's.    

\subsection{Numerical examples}
\def\Tgg1{ $\mathcal{T}_k/N_k \gg 1$}
Eq. (21-23,29) are valid in limit $\mathcal{T}_k/N_k \gg 1$.  It is
of interest to see how their accuracy is degraded at finite values of
these parameters, as well as the sensitivity to other assumptions in
the model.   Table I shows two aspects of the model Hamiltonian and
its reduction to the \Tgg1 limit.  First, one sees that reduction is accurate only
to a 20\% level despite the seeming large value for $  \mathcal{T}_k/N_k =
20$.  The slow convergence can be traced to the {r.m.s.} fluctuation exhibited
in Eq. (\ref{eq:GSD}), dying  off only as $N_k^{-1/2}$.  The table
also demonstrates for $N_2=1$ and $2$  that the transmission probability 
scales quite well with the number of transition states at the same energy, given that 
their couplings are orthogonal and have the same strength. This is
implicit in the reduction to Eq. (\ref{eq:resonances}).
\begin{table}[htb] 
\begin{center} 
\begin{tabular}{|c|cc|} 
\hline 
$N_2$     & \Tgg1 & Eq. (2) \\
\hline
1 & 0.0010 & 0.00079\\
2 & 0.0020 & 0.00148\\
\hline 
\end{tabular}
\caption{Comparison of transmission probability calculated by the trace
formula Eq. (\ref{eq:datta}) and by the \Tgg1 reduction, Eq. (\ref{eq:T_in3-2}).  
The parameter
values in the reservoir Hamiltonians are
$(v_k,v_{k2},\gamma_k,N_k)=
(0.1,0.1,0.1,100)$. The $H_2$ matrix contains $N_2$ transition states
at energies $E_i = 0$, and the partial width of the entrance channel is
$\gamma_{\rm in} = 0.01$.  The ensemble average in Eq. (\ref{eq:datta}) was
carried out with 10000 samples; the statistical uncertainties are of
the order $~1$\%.
} 
\label{exact} 
\end{center} 
\end{table} {

Next we examine the sensitivity to the decay matrix elements.  The
dependence on $\gamma_{\rm in}$ is trivial as it 
contributes quadraticly when it is small compared to other widths.
The exit decay is independent of $\gamma_3$ in the reduced
formula.  This is tested  in 
Table II, varying $\gamma_3$ and keeping the other parameters fixed.
One sees that the dependence is quite flat within the boundaries
$ \rho_3^{-1} <  \gamma_3 <   \sqrt{N_3} v_3$. 
\begin{table}[htb] 
\begin{center} 
\begin{tabular}{|c|c|} 
\hline 
$\gamma_3$  & Eq. (2) \\
\hline
0.05 & 0.00074\\ 
0.1  & 0.00079\\
0.2  & 0.00083\\
0.4  & 0.00085\\
\hline 
\end{tabular}
\caption{Dependence  of transmission probability on the parameter
$\gamma_3$.  The other parameters are the same as in the caption to
Table I for the $N_2=1$ Hamiltonian.
} 
\label{gamma3} 
\end{center} 
\end{table} }

It is also interesting to see how the formula breaks down when 
the condition $\mathcal{T}_k/N_k \gg 1$ is no longer satisfied.  When
the direct decay of the first reservoir becomes small, more
of the probability flux crosses the barrier and $T_{{\rm in},3}$
only competes with elastic scattering.  Table 3 shows a comparison of the analytic
reduction with the full trace evaluation in  Eq. (2).  
\begin{table}[htb] 
\begin{center} 
\begin{tabular}{|c|c|} 
\hline 
$\mathcal{T}_k/N_k$  & R \\
\hline
1 & 2.75\\ 
2  & 2.00\\
6  & 1.45\\
10  & 1.30 \\
20  & 1.25  \\
\hline 
\end{tabular}
\caption{Dependence  of transmission probability on $\mathcal{T}_1/N_1$
varying the parameter $\gamma_1$.  The other parameters are the same as in the caption to
Table I for the $N_2=1$ Hamiltonian.  The column $R$ shows the ratio of the
analytic reduction 
Eq. (\ref{eq:T_in3-2}) 
to value obtained with Eq. (2) taking 10000 samples of
the GOE's. The statistical errors decrease from $~2$\% for the first
entry to $~1$\% for the last one.}
\label{gamma1} 
\end{center} 
\end{table} 
One sees that the analytic reduction becomes quite inaccurate for the
smaller values of $\mathcal{T}_1/N_1$.

\section{summary}

While transition-state theory for decay of quantum complex systems is usually 
derived with a statistical approach, 
we have successfully derived it starting from a matrix Hamiltonian as
is commonly used in configuration-interaction formulations. 
To this end, we considered two reservoirs described by random matrices. 
One of the configurations in the first reservoir undergoes transitions to configurations in 
the second reservoir through bridge configurations between them. 
A potential barrier may exist 
for the bridge configurations. 
This generalizes a model with a single barrier configuration that was
discussed by Weidenm\"uller \cite{weidenmuller2022}.

As in Ref. \cite{weidenmuller2022}, we have shown that the average transmission coefficient from 
the entrance configuration to configurations in the second reservoir can be factorized into a 
product form of the formation and the decay probabilities of transition channels, 
in the limit of $\mathcal{T}_3/N_3\gg1$.
This is also a consequence of the
usual starting point of transition state theory, that once the system passes
the barrier, it never comes back.

If the condition $\mathcal{T}_1/N_1\gg1$ is also satisfied, 
the transmission coefficient is further
simplified to a product of the population probability of the first
reservoir, the transmission coefficient over the barrier, and the
decay probability of the configurations in the second reservoir.  In
that case the transmission coefficient can be expressed in terms of 
Breit-Wigner resonance decays, as has been long
known in nuclear physics and in the field of electron transport.

Transition-state theory is a landmark framework for decays of quantum complex systems, but 
conditions for transition-state theory to work have not yet been well clarified. 
The microscopic derivation based on the random matrix approach shown in this paper 
provides a necessary 
condition for transition-state theory to work. 
Such consideration would be important in the decay of 
complex systems at energies close to barrier tops.

\begin{acknowledgments}
We thank Hans Weidenm\"uller for useful discussions. 
This work was supported in part by
JSPS KAKENHI Grant Numbers JP19K03861 and JP23K03414.
\end{acknowledgments}

\appendix

\section{the Green's function for a block-tridiagonal Hamiltonian}

We invert the matrix 
\begin{equation}
H-i\tilde\Gamma_{\rm in}/2-i\tilde\Gamma_1/2-i\tilde\Gamma_3/2-E=
\left(
\begin{matrix} 
\tilde{H}_1 & V_{12} & 0 \cr
V_{12}^T & \tilde{H}_2 & V_{23} \cr
0 & V_{23}^T & \tilde{H}_3
\end{matrix}
\right),
\end{equation}
where $\tilde{H}_i$ are defined as 
\begin{eqnarray}
\tilde{H}_1&\equiv& H_1-i\Gamma_{\rm in}/2-i\Gamma_1/2-E, \\ 
\tilde{H}_2&\equiv& H_2-E, \\
\tilde{H}_3&\equiv& H_1-i\Gamma_3/2-E. 
 \end{eqnarray}
The Green's function (\ref{eq:G}) satisfies the relation, 
\begin{equation}
\left(
\begin{matrix} 
\tilde{H}_1 & V_{12} & 0 \cr
V_{12}^T & \tilde{H}_2 & V_{32}^T \cr
0 & V_{32} & \tilde{H}_3
\end{matrix}
\right)
\left(
\begin{matrix} 
G_{11} & G_{12} & G_{13} \cr
G_{21} & G_{22} & G_{23} \cr
G_{31} & G_{32} & G_{33} 
\label{eq:green}
\end{matrix}
\right)
=
\left(
\begin{matrix} 
1 & 0 & 0 \cr
0 & 1 & 0 \cr
0 & 0 & 1 
\end{matrix}
\right), 
\end{equation}
from which one finds 
\begin{eqnarray}
&&\tilde{H}_1G_{13}+V_{12}G_{32}^T=0, \label{eq:G1} \\
&&V_{12}^TG_{13}+\tilde{H}_2G_{23}+V_{23}G_{33}=0, \label{eq:G2} \\
&&V_{32}G_{23}+\tilde{H}_3G_{33}=1. \label{eq:G3}
\end{eqnarray}
From Eqs. (\ref{eq:G1}) and (\ref{eq:G3}), $G_{13}$ and $G_{33}$ read
\begin{equation}
G_{13}=-\tilde{H}_1^{-1}V_{12}G_{23},
\label{eq:G13-2}
\end{equation}
and
\begin{equation}
G_{33}=\tilde{H}_3^{-1}-\tilde{H}_3^{-1}V_{32}G_{23},
\end{equation}
respectively. 
Substituting these into Eq. (\ref{eq:G2}), one obtains 
\begin{equation}
G_{23}=-(\tilde{H}_2-V_{12}^T\tilde{H}_1^{-1}V_{12}-V_{32}^T\tilde{H}_3^{-1}V_{32})^{-1}
V_{32}^T\tilde{H}_3^{-1}.
\label{eq:G23}
\end{equation}
Combining Eqs. (\ref{eq:G13-2}) and (\ref{eq:G23}), 
one finally obtains Eq. (\ref{eq:G13}). 

Following a similar procedure, one can also derive
\begin{equation}
    G_{11}=G_1+G_1V_{12}G_2V_{12}^TG_1. 
\end{equation}

\section{Ensemble average of $VG\tilde\Gamma G^\dagger V^T$}

In this Appendix, we evaluate the ensemble average of a matrix
$VG\tilde\Gamma G^\dagger V^T$, where the elements of $V$ are
Gaussian-distributed random numbers with $\langle v_{ij}^2\rangle = v^2$,
$\tilde \Gamma $ is a constant times the unit matrix
$\tilde\Gamma_{ij}=\gamma\delta_{i,j}$,
and $G$ is the Green's function $G=(H-i\tilde\Gamma/2)^{-1}$. 
Here  $H$ is a sample of the $N\times N$
GOE with a level density $\rho_0$ in the center of its spectrum.  
We follow Refs.  \cite{bertsch-hagino2021,hagino-bertsch2021}
to carry out the ensemble averaging.  We first 
express the elements of the Green's function as
\begin{equation}
G_{ij}=\sum_\lambda\frac{\phi^{\lambda}_{i}\phi^{\lambda}_{j}}{E_\lambda-i\gamma/2}, 
\end{equation}
where $E_\lambda$ are the eigenvalues of the Hamiltonian $H$ and $\phi^\lambda$ 
are the corresponding eigenfunctions. 

The $ij$ element of $VG\tilde\Gamma G^\dagger V^T$ then reads,
\be
(VG\tilde\Gamma G^\dagger V^T)_{ij} = \sum_{i'j'mm'\lambda\lambda'} \frac{V_{ii'}
\phi^{\lambda}_{i'}\phi^{\lambda}_{m}\tilde{\Gamma}_{mm'}\phi^{\lambda'}_{m'}\phi^{\lambda'}_{j'}V_{j'j}}
{(E_\lambda-i\gamma/2)(E_{\lambda'}+i\gamma/2)}. 
\label{eq:Veff}
\ee
The sum over $m'$ is evaluated as
\be
\sum_{m'} \tilde{\Gamma}_{mm'}\phi^{\lambda'}_{m'} = \gamma \phi^{\lambda'}_m, 
\ee
since $\tilde{\Gamma}$ is assumed to be proportional to the unit matrix.
Next the orthogonality of the eigenvectors $\lambda$ and $\lambda'$  permits the
sum over $\lambda'$ to be dropped with replacement $\lambda'$ by
$\lambda$. Then Eq. (\ref{eq:Veff}) reduces to
\be
(VG\tilde\Gamma G^\dagger V^T)_{ij}  = \gamma\sum_{i' j'}\sum_\lambda 
\left(\frac{V_{ii'}V_{jj'}\phi^\lambda_{i'}\phi^\lambda_{j'}}{E_\lambda^2 +
\gamma^2/4}  \right).
\label{eq:b4}
\ee  
Next we take the ensemble average of the factor in parentheses.  One of the
properties of the GOE is that fluctuations about the average level densities
are small, so the ensemble averages of the numerator and denominator are
uncorrelated.  The denominator average is \cite{bertsch-hagino2021,hagino-bertsch2021} 
\be
\left\langle\sum_\lambda \frac{1}{E_\lambda^2+\gamma^2/4}\right\rangle
= 2 \pi \frac{\rho_0} {\gamma}.
\ee
For the numerator, we first notice that the dot products 
$\vec{V}_i\cdot \vec{\phi}_\lambda\equiv \sum_{i'}V_{ii'}\phi_{i'}^\lambda$ are Gaussian distributed 
with $\vec{V}_i\cdot \vec{\phi}_\lambda=v\,r_{i\lambda}$, where $r_{i\lambda}$ is a random number 
satisfying $\langle r_{i\lambda}\rangle=0$ and $\langle r_{i\lambda}r_{i'\lambda'}\rangle=\delta_{i,i'}\delta_{\lambda\lambda'}$. 
One thus obtains 
\begin{equation}
\langle(\vec{V}_i\cdot\vec\phi_\lambda)(\vec{V}_j\cdot\vec\phi_{\lambda})\rangle
=v^2\delta_{i,j}, 
\end{equation}
which leads to 
\be
    \left\langle (VG\tilde\Gamma G^\dagger V^T)_{ij}\right\rangle
    =2 \pi v^2 \rho_0\,\delta_{i,j}. 
\label{eq:VGGV}
\ee

We also need the ensemble average of $VG\tilde\Gamma_{\rm in} G^\dagger V^T$ with  
$G=(H-i\tilde\Gamma/2-i\tilde\Gamma_{\rm in}/2)^{-1}$, where $\tilde\Gamma_{\rm in}$ is 
given by Eq. (\ref{eq:Gin2}).  
In this case, the sum over $m$ is restricted to a single state in  Eq. (\ref{eq:Veff}). 
Due to the invariance of the averages under unitary transformations,
the sum over $m,m'$ becomes
$\langle
\phi_{\lambda, {\rm in}}\phi_{\lambda',\rm{in}}
\rangle=\delta_{\lambda,\lambda'}/N$.
The final result is
\begin{equation}
    \left\langle (VG\tilde\Gamma_{\rm in} G^\dagger V^T)_{ij}\right\rangle
    =\frac{\Gamma_{\rm in}}{N \gamma}2 \pi v^2 \rho_0\,\delta_{i,j}. 
\label{eq:VG_inGV}
\end{equation}.


\begin{thebibliography}{99} 

\bibitem{hanggi1990}
P. H\"anggi, P. Talkner, and M. Borkovec, 
Rev. Mod. Phys. {\bf 62}, 251 (1990). 

\bibitem{tr96} D.G. Truhlar, B.C. Garrett, and S.J. Klippenstein,
J. Phys. Chem. {\bf 100} 12771 (1996).

\bibitem{weiss2012}
U. Weiss, {\it Quantum Dissipative Systems 4th edition}, (World Scientific, 
Singapore, 2012). 

\bibitem{BW39}N. Bohr and J.A. Wheeler,   
Phys. Rev.  {\bf 56}, 426 (1939).

\bibitem{MR51}R.A. Marcus and O.K. Rice,  
J. Phys. and Colloid Chem.  {\bf 55}, 894 (1951).

\bibitem{weidenmuller2009}
H.A. Weidenm\"uller, Rev. Mod. Phys. {\bf 81}, 539 (2009).

\bibitem{bertsch-hagino2023}
G.F. Bertsch and K. Hagino, Phys. Rev. C{\bf 107}, 044615 (2023). 

\bibitem{polik1990}
W.F. Polik, D.R. Guyer, W.H. Miller, and C.B. Moore, J. Chem. Phys. {\bf 92}, 3471 (1990).

\bibitem{miller1990}
W.H. Miller, R. Hernandez, C.B. Moore, and W.F. Polik, J. Chem. Phys. {\bf 93}, 5657 (1990).

\bibitem{hernandez1993}
R. Hernandez, W.H. Miller, C.B. Moore, and W.F. Polik, J. Chem. Phys. {\bf 99}, 950 (1993).


\bibitem{weidenmuller2022}
H.A. Weidenm\"uller, Phys. Rev. E{\bf 105}, 044143 (2022). 

\bibitem{weidenmuller2023}
H.A. Weidenm\"uller, arXiv:2311.08030.

\bibitem{alhassid2000}
Y. Alhassid, Rev. Mod. Phys. {\bf 72}, 895 (2000). 


\bibitem{da01} P.S. Damle, A.W. Ghosh, and S. Datta, Phys. Rev. B {\bf 64}, 
201403 (2001).

\bibitem{da95} S. Datta, {\it Electronic Transport in Mesoscopic Systems}, 
(Cambridge University Press, Cambridge, 1995), Eq. (3.5.20).


\bibitem{mil93}
W.H. Miller, Acc. Chem. Res. {\bf 26}, 174 (1993). 

\bibitem{ha08} H. Haug and A-P. Jauho, {\it 
Quantum kinetics in transport and optics of semiconductors},
(Spring, Berlin, 2008), Eq. (12.37).



\bibitem{al21} Y.~Alhassid, G.F.~Bertsch, and P.~Fanto, Ann. Phys. {\bf 424} 
168381 (2021). 

\bibitem{meir1992}
Y. Meir and N.S. Wingreen, Phys. Rev. Lett. {\bf 68}, 2512 (1992), Eq. (7).

\bibitem{buttiker1986}
M. B\"uttiker, Phys. Rev. Lett. {\bf 57}, 1761 (1986).

\bibitem{venugopal2003}
R. Venugopal, M. Paulsson, S. Goasguen, S. Datta, 
and M.S. Lundstrom, J. of App. Phys. {\bf 93}, 5613 (2003).

\bibitem{datta2005}
S. Datta, {\it Quantum Transport: Atom to Transistor} (Cambridge University 
Press, Cambridge, 2005). 

\bibitem{hagino-bertsch2021}
K. Hagino and G.F. Bertsch, Phys. Rev. E{\bf 104}, L052104 (2021). 

\bibitem{Lo2000} O.I. Lobkis, R.L. Weaver, and I. Rozkhov, J. Sound
Vib. {\bf 237}, 281, (2000).
\bibitem{Fe2020} S.B. Fedeli and Y.V. Fyodorov, J. Phys. A: Math.
Theor. {\bf 53} 165701 (2020).


\bibitem{bertsch-hagino2021}
G.F. Bertsch and K. Hagino, J. Phys. Soc. Jpn. {\bf 90}, 114005 (2021). 


\bibitem{bertsch2014}
G.F. Bertsch, arXiv: 1407.1899 [nucl-th]. 

\end{thebibliography}
\end{document}